\documentstyle[aps,preprint,epsf,floats]{revtex}

\renewcommand{\slash}[1]{#1 \hspace{-0.55em} / }

\def\bq{\begin{eqnarray}}
\def\eq{\end{eqnarray}}
\def\be{\begin{eqnarray}}
\def\ee{\end{eqnarray}}
\def\ben{\begin{enumerate}}\def\een{\end{enumerate}}

\def\roughly#1{\mathrel{\raise.3ex\hbox{$#1$\kern-.75em
\lower1ex\hbox{$\sim$}}}}

\begin{document}

\preprint{\hfill\parbox[b]{0.3\hsize}
{ FTUV-03-708; IFIC-03-35 }}

\def\bra{\langle }
\def\ket{\rangle }

\title{
Generalized Parton Distributions
and composite constituent quarks
\thanks{
Supported in part by GV-GRUPOS03/094, 
MCYT-FIS2004-05616-C02-01 and by MIUR through the funds COFIN03}
}

\author{
Sergio Scopetta}
\address{
Dipartimento di Fisica, Universit\`a degli Studi
di Perugia, via A. Pascoli
06100 Perugia, Italy
\\
and INFN, sezione di Perugia}

\author{Vicente Vento}
\address{
Departament de Fisica Te\`orica,
Universitat de Val\`encia, 46100 Burjassot (Val\`encia), Spain
\\
and Institut de F\'{\i}sica Corpuscular,
Consejo Superior de Investigaciones Cient\'{\i}ficas}

\maketitle

\begin{abstract}
An approach is proposed to calculate Generalized Parton Distributions
(GPDs) in a Constituent Quark Model (CQM) scenario, considering
the constituent quarks as complex systems.
The GPDs are obtained 
from the wave functions of the 
non relativistic CQM of Isgur and Karl,
convoluted with the GPDs of the constituent quarks
themselves. The latter are modelled by using 
the structure functions
of the constituent quark, the double distribution representation
of GPDs, and a recently proposed phenomenological 
constituent quark form factor.
The present approach permits to access a kinematical
range corresponding
to both the DGLAP and the ERBL regions, 
for small values 
of the momentum transfer
%, $\Delta^2$
%($\vec \Delta ^2 \ll m^2$, where $m$ is the constituent
%quark mass) 
and of the skewedness parameter.
%, $\xi$.
In this kinematical region,
the cross sections relevant
to deeply virtual Compton scattering could be estimated
by using the obtained GPDs.
As an example,
the leading twist, unpolarized
GPD $H$ has been calculated. 
Its general relations with 
the non relativistic definition 
of the electric form factor and with the leading
twist unpolarized quark density are consistently recovered
from our expressions.
Further natural applications of the proposed approach
%and the benefits one would get from a relativistic description 
are addressed.
\end{abstract}
\pacs{12.39-x, 13.60.Hb, 13.88+e}
%\noindent{\small$\dagger$

\section{Introduction}

Generalized Parton Distributions (GPDs) \cite{first} 
parametrize the non-perturbative hadron structure
in hard exclusive 
processes
(for comprehensive reviews, 
see, e.g.,  \cite{jig,rag,pog,dpr,kroll,freufreu}).
The measurement of
GPDs would provide information which is
usually encoded in both the elastic form factors
and 
the usual Parton Distribution Functions (PDFs)
and, 
at the same time, it would represent 
a unique way to access several 
crucial features of the structure of the nucleon
\cite{radnew,ji1}. 
By measuring GPDs,
a test of the Angular Momentum Sum Rule
of the proton \cite{jaffe} could be
achieved for the first time, determining
the quark orbital angular momentum contribution to the proton spin
\cite{ji1,jiprd}.

Besides, the possibility of obtaining, by means
of GPDs measurements, information on the structure of the proton
in the impact parameter \cite{burk,diehl} 
and position \cite{piral,bmo,jil}
spaces, is being presently discussed. 

Therefore,
relevant experimental efforts to measure GPDs,
by means of exclusive electron Deep Inelastic Scattering
(DIS) off the proton, are likely to
take place in the next few years \cite{ceb,hera,esop}.

In this scenario, it becomes urgent to produce 
theoretical predictions for the behavior of these quantities.
Several calculations have been already performed by using different 
descriptions 
of hadron structure: bag models \cite{meln,vv}, soliton models
\cite{pog,goe1}, light-front \cite{mill}
and Bethe Salpeter approaches \cite{lukas},
phenomenological estimates 
based on parametrizations of PDFs
\cite{freund,rad1}.
Besides, an impressive effort has been devoted to study
the perturbative QCD
evolution \cite{scha2,evo} of GPDs, and the GPDs at twist three
accuracy \cite{scha1}.

Recently, calculations have been performed also
in Constituent Quark Models (CQM) \cite{epj,bpt}.
The CQM has a long story of successful
predictions in low energy studies of the electromagnetic
structure of the nucleon.
In the high energy sector,
in order to compare model predictions
with data taken in DIS experiments, 
one has to evolve, according to perturbative QCD, the leading twist
component of the physical structure functions obtained
at the low momentum scale associated with the model,
the so called ``hadronic scale'', $\mu_0^2$.

Such a procedure, already addressed in \cite{pape,jaro}, 
has proven
successful in describing the gross features of 
standard PDFs 
by using different CQM
(see, e.g., \cite{trvv}).
Similar expectations motivated the study of GPDs
in Ref. \cite{epj}.
In that paper, a simple formalism has been proposed to calculate
the quark contribution to 
GPDs from any non relativistic or relativized model
and, as an illustration, results obtained in the
Isgur and Karl model \cite{ik} have been evolved
from $\mu_0^2$ up to DIS scales, to NLO accuracy.
In Ref. \cite{bpt} the same quark contribution to GPDs has been
evaluated, at $\mu_0^2$,  using the 
overlap representation of GPDs \cite{kroll}
in light-front dynamics,
along the lines developed in \cite{pietro}.

In here, the procedure of Ref. \cite{epj} is extended
and generalized.
As a matter of fact, the approach of Ref. \cite{epj}, when
applied in the standard
forward case, has been proven to
reproduce the gross features of PDFs
\cite{trvv} but,
in order to achieve a better agreement with data, it
has to be improved.
In a series of papers, it has been shown that
unpolarized
\cite{scopetta1} and polarized \cite{scopetta2} DIS data
are consistent with a low energy scenario,
dominated by complex constituent quarks
inside the nucleon, defined through a scheme suggested
by Altarelli,
Cabibbo, Maiani and Petronzio (ACMP) \cite{acmp},
updated with modern phenomenological
information. The same idea has been recently applied to demonstrate
the evidence of complex objects inside the nucleon
\cite{psr}, analyzing intermediate energy data of electron scattering off 
the proton. Besides,
a similar scenario has been extensively used by other
groups, starting form the concept of ``valon'',
introduced more than twenty years ago \cite{hwa}
(for recent developments, see \cite{hwa1}).

We here generalize our description of the forward case \cite{scopetta1}
to the calculation scheme of Ref. \cite{epj}, 
in order to obtain 
more realistic
predictions for the GPDs and, at the same time, explore
kinematical regions not accessible before.

In particular, the evaluation
of the sea quark contribution becomes possible, so that
GPDs could be calculated, in principle, in their full range of definition.
Such an achievement would permit to estimate
the cross-sections which are relevant for actual GPDs
measurements, providing us with an important
tool for planning 
future experiments.
Actually, as it will be shown, the proposed approach
will be applied here in a Non Relativistic (NR) framework,
which allows one to evaluate the GPDs only 
for small values 
of the 4-momentum transfer, $\Delta^2$
(corresponding to $\vec \Delta ^2 \ll m^2$, where $m$ is the constituent
quark mass) and small values also for
the skewedness parameter, $\xi$.
The full kinematical range of definition of GPDs will be
studied in a followup, introducing relativity
in the scheme.

The paper is structured as follows.
After the definition of the main quantities
of interest, an Impulse Approximation (IA) convolution formula for
the current quark GPDs in terms of the constituent quark
off-diagonal momentum distributions and constituent quark
GPDs is derived in the third section. 
Then, the constituent quark GPDs are built in the fourth section,
according to ACMP philosophy and 
using the Double Distribution (DD's) representation 
\cite{rag,rad1,radd} of the GPDs.
In the fifth section, as an illustration,
results obtained by using CQM wave functions of the IK model
and the obtained constituent quark GPDs will be shown.
Conclusions will be drawn in the last section.

\section{Quark Model Calculations of GPDs}

We are interested in hard exclusive processes.
The absorption of a high-energy
virtual photon by a quark in a 
hadron target is followed by
the emission of a particle to be later
detected; finally, the interacting quark
is reabsorbed back into the recoiling hadron.
If the emitted and detected particle is, for example, a real photon,
the so called Deeply Virtual Compton Scattering
process takes place  \cite{radnew,ji1,jiprd}.
We adopt here the formalism used in Ref. \cite{jig}.
Let us think to a nucleon target, with initial (final)
momentum and helicity $P(P')$ and $s(s')$, 
respectively. 
The GPDs $H_q(x,\xi,\Delta^2)$ and
$E_q(x,\xi,\Delta^2)$
are defined through the expression
\begin{eqnarray}
\label{eq1}
F^q_{s's}(x,\xi,\Delta^2) & = &
{1 \over 2} \int {d \lambda \over 2 \pi} e^{i \lambda x}
\bra P' s' | \, \bar \psi_q \left(- {\lambda n \over 2}\right)
\slash{n} \, \psi_q \left({\lambda n \over 2} \right) | P s \ket  =  
\nonumber
\\
& = & H_q(x,\xi,\Delta^2) {1 \over 2 }\bar U(P',s') 
\slash{n} U(P,s) + 
E_q(x,\xi,\Delta^2) {1 \over 2} \bar U(P',s') 
{i \sigma^{\mu \nu} n_\mu \Delta_\nu \over 2M} U(P,s)~,
\end{eqnarray}
\\
where 
$\Delta=P^\prime -P$
is the 4-momentum transfer to the nucleon,
$\psi_q$ is the quark field and M is the nucleon mass.
It is convenient to work in
a system of coordinates where
the photon 4-momentum, $q^\mu=(q_0,\vec q)$, and $\bar P=(P+P')/2$ 
are collinear along $z$.
The $\xi$ variable in the arguments of the GPDs 
is the so called ``skewedness'', parametrizing
the asymmetry of the process. It is defined
by the relation $\xi = - n \cdot \Delta/2$,
where $n$
is a light-like 4-vector
satisfying the condition $n \cdot \bar P = 1$.
%Besides, one has $t=\Delta^2=
%\Delta_0^2 - \vec{\Delta}^2$. 
As explained in \cite{ji1,jiprd},
GPDs describe the amplitude for finding a quark with momentum fraction
$~~x+\xi$ (in the Infinite Momentum Frame) in a nucleon 
with momentum $(1+\xi) \bar P$
and replacing it back into
the nucleon with a momentum transfer $\Delta$.
Besides, when the quark longitudinal momentum fraction 
$x$ of the average nucleon momentum $\bar P$
is less than $-\xi$, GPDs describe antiquarks;
when it is larger than $\xi$, they describe quarks;
when it is between $-\xi$ and $\xi$, they describe 
$q \bar q$ pairs.
The first and second case  are commonly referred to as
DGLAP region and the third as ERBL region \cite{jig},
following the pattern of evolution in the factorization scale.
%The region $|x| \geq \xi/2$ is often called the DGLAP region,
%since the $Q^2$ evolution of GPDs is governed there
%by the DGLAP equations of perturbative QCD \cite{dglap};
%the region $|x| \leq \xi/2$ is called the ERBL region,
%because there the $Q^2$ evolution is ERBL-like \cite{erbl}.
One should keep in mind that, besides the variables
$x,\xi$ and $\Delta^2$ explicitly shown, GPDs depend,
as the standard PDFs, on the momentum scale $Q^2$ at which they are
measured or calculated. For an easy presentation,
this latter dependence will be 
omitted in the rest of the paper, unless specifically needed.
The values of $\xi$ which are possible for a given value of
$\Delta^2$ are:
\bq
0 \le \xi \le \sqrt{- \Delta^2}/\sqrt{4 M^2-\Delta^2}~.
\label{xim}
\eq
The well known natural constraints of $H_q(x,\xi,\Delta^2)$ are: 

i) the so called
``forward'' or ``diagonal'' limit, 
$P^\prime=P$, i.e., $\Delta^2=\xi=0$, where one 
recovers the usual PDFs
\bq
H_q(x,0,0)=q(x)~;
\label{i)}
\eq

ii)
the integration over $x$, yielding the contribution
of the quark of flavor $q$ to the Dirac 
form factor (f.f.) of the target:
\bq
\int dx H_q(x,\xi,\Delta^2) = F_1^q(\Delta^2)~;
\label{ii)}
\eq

iii) the polynomiality property \cite{jig},
involving higher moments of GPDs, according to which
the $x$-integrals of $x^nH^q$ and of $x^nE^q$
are polynomials in $\xi$ of order $n+1$.

In \cite{epj},
the IA expression
for 
$H_q(x,\xi,\Delta^2)$, suitable to perform CQM
calculations, has been obtained.

Now, it will be shown that the same basic 
formula can be derived
as the
NR reduction of the definition (\ref{eq1}) of GPDs,
analyzed initially in the non-covariant
framework of light-cone quantization, involving partons on their mass shell. 
%Such a procedure is helpful if one wants to assign helicity to them.

Using the light-cone spinor definitions as given
in the appendix B of \cite{dpr}, 
and defining: 
$k^+=(k_0+k_3)/\sqrt{2}, \vec k_\perp = (k_1,k_2)$,
for the light-cone helicity
combination $s's={1 \over 2} {1 \over 2} =++$ one obtains
\begin{eqnarray}
F^q_{++}(x,\xi,\Delta^2) = {\sqrt{1 - \xi^2} } 
% \over \sqrt{2}} 
H_q(x,\xi,\Delta^2) - {\xi^2 \over 
\sqrt{1 - \xi^2}} E_q(x,\xi,\Delta^2)~,
\end{eqnarray} 
so that, for $\xi^2 \ll 1$:

\begin{eqnarray}
F^q_{++}(x,\xi,\Delta^2)  = 
{H_q(x,\xi,\Delta^2)} 
%\over \sqrt{2}} 
- \xi^2 
\left( {1 \over 2} H_q(x,\xi,\Delta^2)  
%\over 2 \sqrt{2} } 
+ E_q(x,\xi,\Delta^2)  \right)
+ O(\xi^4)~,
\end{eqnarray} 

i.e.

\begin{eqnarray}
F^q_{++} (x,\xi,\Delta^2) = H_q (x,\xi,\Delta^2) 
%\over \sqrt{2} }~.
+ O(\xi^2)~. 
\label{hf++}
\end{eqnarray} 

The reader should be aware that the so called ``Munich Symmetry''
for double distributions excludes $O(\xi)$ contributions
to GPDs \cite{mun},
so that the accuracy of the above equation
is worse than it reads.

According to the latter equation,
in order to obtain the GPD $H_q(x,\xi,\Delta^2)$ 
for $\xi^2 \ll 1$ one has to evaluate $F_{++}^q$, starting from its definition,
Eq (\ref{eq1}). In the l.h.s. of the latter,
using light-cone quantized quark fields,
whose creation
and annihilation operators,
$b^\dag(k)$ and
$b(k)$, obey the commutation relation 
\bq
\{ b(k'),b^\dag(k)\} = (2 \pi)^3 2 k^+ \delta(k'^+ - k^+)
\delta^2(\vec k_\perp - \vec k_\perp')~,
\eq
and using properly normalized light-cone states
\bq
\bra P'  | P \ket = (2 \pi)^3 2 P^+ \delta(P'^+ - P^+)
\delta^2(\vec P_\perp - \vec P_\perp')~,
\eq
one obtains, for $x > \xi$ 
\cite{jig}
\bq
F^q_{++} (x,\xi,\Delta^2)
= {1 \over 2 \bar P^+ V} \int 
{d^2 k_\perp \over 2 \sqrt{ |x^2 - \xi^2|} (2 \pi)^3 } \bra 
b^\dag (k + \Delta) b(k)  \ket ~,
\label{fqpp}
\eq  
where
\bq
\bra b^\dag(k') b(k) \ket= \sum_\lambda \bra P'+ | 
b_\lambda^\dag((x-\xi)\bar{P}^+, {\vec k_\perp'} )
b_\lambda((x+\xi)\bar{P}^+, {\vec k_\perp})| P+ \ket~,
\eq
and $V$ is a volume factor.
We are interested here in the $x > \xi$ region,
since we want to obtain only the quark contribution to $H$,
the only one which can be evaluated in a CQM with
three valence, point-like quarks.
 
Eq. (\ref{fqpp}) can be written:
\bq
F^q_{++} (x,\xi,\Delta^2)
& = & {1 \over 2 \bar P^+ V } \int 
{d^2 k_\perp \over 2 \sqrt{ |x^2 - \xi^2|} (2 \pi)^3}  \bra 
b^\dag (k + \Delta) b(k)  \ket =
\nonumber
\\
& = &
{1 \over 2 \bar P^+ V} \int 
{d^2 k_\perp d k^+ \over 2 \sqrt{ |x^2 - \xi^2|} (2 \pi)^3 } 
\delta(k^+ - (x+\xi)\bar P^+)
\bra b^\dag (k + \Delta) b(k)  \ket =
\nonumber
\\
& = &
{1 \over 2 \bar P^+ V} \int 
{d^2k_\perp d k^+ \over 2 \sqrt{ (x+\xi)\bar P^+
(x-\xi) \bar P^+} (2 \pi)^3 } 
\bar P^+ \times \nonumber
\\
& \times &
\delta(k^+ - (x+\xi)\bar P^+)
\bra b^\dag (k + \Delta) b(k)  \ket =
\nonumber
\\
& = &
{1 \over 2 \bar P^+ V} \int 
{d^2 k_\perp d k^+ \over 2 \sqrt{ k^+
k'^+} (2 \pi)^3 } 
\delta \left( {k^+ \over \bar P^+} - (x+\xi) \right)
\bra b^\dag (k + \Delta) b(k)  \ket~. 
\label{fqpp1}
\eq 

In a NR framework,
states and creation and annihilation operators 
have to be normalized according to
\bq
\bra \vec P' | \vec P \ket = (2 \pi)^3 
\delta(P'^+ - P^+)
\delta(\vec P'_\perp - \vec P_\perp)~
\eq
and
\bq
\{ b(k'^+,\vec k_\perp),b^\dag (k^+, \vec k'_\perp)\} = (2 \pi)^3 
\delta(k'^+ - k^+)
\delta(\vec k'_\perp - \vec k_\perp)~
\eq
respectively. As a consequence, 
in order to perform a NR reduction of Eq. (\ref{fqpp1}),
one has to consider that \cite{muld}:
\bq
| P \ket \rightarrow \sqrt{2 P^+}|\vec P \ket~,
\eq
\bq
b(k) \rightarrow \sqrt{2 k^+}b(k^+,\vec k_\perp)~,
\eq
so that in Eq. (\ref{fqpp1}), in terms of the
new states and fields,
one has to perform the substitution
\bq
\bra b^\dag (k + \Delta) b(k) \ket 
& = & 
\sum_\lambda \bra P'+ | 
b_\lambda^\dag((x-\xi)\bar{P}^+, \vec k_\perp + \vec \Delta_\perp)
b_\lambda((x+\xi)\bar{P}^+, \vec k_\perp )| P + \ket  \rightarrow 
\nonumber
\\
& \rightarrow &
2 \sqrt{1 - \xi^2}\bar P^+ 
\sqrt{2 k'^+ 2 k^+}
\sum_\lambda 
\bra \vec P' | 
b_\lambda^\dag(k^+ + \Delta^+, \vec k_\perp + \vec \Delta_\perp )
b_\lambda(k^+, \vec k_\perp )| \vec P \ket~,
\label{newnorm}
\eq
where use has been made of the relation
$2 \sqrt{1 - \xi^2}\bar P^+ = \sqrt{2 P^+} \sqrt{2 P'^+}$.

Now, inserting Eq. (\ref{newnorm}) in Eq. (\ref{fqpp1}),
one gets
\bq
F_{++}^q(x,\xi,\Delta^2)
& = &
{ 1 \over V} \int 
{d^2 k_\perp d k^+ \over 2 \sqrt{ k^+
k'^+} (2 \pi)^3 } 
\delta \left( {k^+ \over \bar P^+} - (x+\xi) \right)
\times
\nonumber
\\
& \times &
\sqrt{2 k'^+ 2 k^+}
\sum_\lambda 
\bra \vec P' | 
b_\lambda^\dag(k^+ + \Delta^+, \vec k_\perp + \vec \Delta_\perp )
b_\lambda(k^+, \vec k_\perp )| \vec P \ket + O(\xi^2)=
\nonumber
\\
& = & 
{1 \over V} \int 
{d^2 k_\perp d k^+ \over (2 \pi)^3 } 
\delta \left( {k^+ \over \bar P^+} - (x+\xi) \right)
\times
\nonumber
\\
& \times &
\sum_\lambda 
\bra \vec P' | 
b_\lambda^\dag(k^+ + \Delta^+, \vec k_\perp + \vec \Delta_\perp)
b_\lambda(k^+, \vec k_\perp )| \vec P \ket + O(\xi^2)~.
\label{intm}
\eq 

Now, since the constituent quarks with mass $m$ are taken to be on shell,
so that $k_0 = \sqrt{ \vec k^2 + m^2}$, one has
\bq
d^2 k_\perp d k^+ \rightarrow  
\left( { k^+ \over k_0} \right) d \vec k~,
\eq
so that, from Eq. (\ref{intm}) 
and Eq. (\ref{hf++}), one gets:
\bq
H_q(x,\xi,\Delta^2)
& = &
\int 
{d \vec k } \,\,
\delta \left( {k^+ \over \bar P^+} - (x+\xi) \right)
\times
\nonumber
\\
& \times &
\left [ 
{ 1 \over (2 \pi)^3 V}
{k^+ \over k_0}
\sum_\lambda 
\bra \vec P' | 
b_\lambda^\dag(k^+ + \Delta^+, \vec k_\perp + \vec \Delta_\perp)
b_\lambda(k^+, \vec k_\perp )| \vec P \ket \right] + O(\xi^2) =
\label{int_1}
\\
& = &
\int  d \vec k \,\, 
%\left[ 1 + O \left( {\vec k^2 \over m^2} \right) \right] 
\delta \left ( {k^+ \over \bar P^+} - (x+\xi) \right )  
%\times
%\nonumber
%\\
%& \times &
\left[\,n_q ( \vec k, \vec k + \vec \Delta)
+ O \left( {\vec k ^2 \over m^2}, {(\vec k + \vec \Delta)^2 \over m^2} 
\right)  \, \right] + O(\xi^2)~.
\label{inter}
\eq
In the last step, 
the definition of (non-diagonal) momentum distribution,
$\,n_q ( \vec k, \vec k + \vec \Delta)$,
has been used, together with the
fact that a NR momentum distribution
describes the probability of finding a constituent
of momentum $\vec k$ in a given system
up to terms
of order ${\vec k^2 \over m^2}$ \cite{fs}.

Summarizing, we find that, in a NR CQM, the GPD $H_q(x,\xi,\Delta^2)$
can be calculated, for $\xi^2 \ll 1$, $\vec k^2 \ll m^2$ and
$(\vec k + \vec \Delta)^2 \ll m^2$ (which means, in turn, 
$\vec \Delta^2 \ll m^2$), through the following expression:
\begin{eqnarray}
H_q(x,\xi,\Delta^2) =
\int  d \vec k \,
\delta \left( {k^+ \over \bar P^+} - (x+\xi) \right)
\,n_q ( \vec k, \vec k + \vec \Delta)
+ O \left( \xi^2, {\vec k ^2 \over m^2}, {\vec \Delta^2 \over m^2} 
\,\right) 
~.
\label{usu}
\end{eqnarray}
The above equation, corresponding to
Eq. (8) in \cite{epj}, 
permits the calculation of 
$H_q(x,\xi,\Delta^2)$ in any CQM, and
it naturally verifies some of the properties
of GPDs.
In fact,
the unpolarized quark density, $q(x)$, 
as obtained  by analyzing,
in IA, DIS off the nucleon
(see, e.g., \cite{muld}),
assuming that the interacting quark is on-shell,
is recovered in the forward limit,
where $\Delta^2=\xi=0$:
\begin{eqnarray}
q(x) = H_q(x,0,0)=
\int 
d \vec k 
\, n_q(\vec k) \, \delta \left ( x - {k^+ \over \bar P^+}
\right )~,
\label{hf}
\end{eqnarray}
so that the constraint Eq. (\ref{i)}) is fulfilled.
In the above equation, $n_q(\vec k)$ is the momentum distribution
of the quarks in the nucleon:
\begin{eqnarray}
n_q(\vec k) =  \int e^{i \vec k \cdot ( \vec r - \vec r')}
\rho_q(\vec r, \vec r') d \vec r
d \vec r'~.
%\nonumber 
\end{eqnarray}

Besides, integrating Eq. (\ref{usu}) over $x$, 
one obtains
\begin{eqnarray}
\int d x H_q(x,\xi, \Delta^2) =
\int d \vec r e^{i \vec \Delta \cdot \vec r} \rho_q(\vec r)
~,
\nonumber
\end{eqnarray}
where $\rho_q(\vec r) = \lim_{\vec r' \rightarrow \vec r}
\rho_q(\vec r', \vec r)$ is the contribution 
of the quark $q$ to the charge density.
The r.h.s. of the above 
equation gives the IA definition of the charge f.f. 
\begin{eqnarray}
\int d \vec r e^{i \vec \Delta \cdot \vec r} \rho_q(\vec r)=
F^q(\Delta^2)
~,
\label{ffr}
\end{eqnarray}
so that, recalling that
$F^q(\Delta^2)$ coincides
with
the non relativistic limit of the Dirac f.f. $F_1^q(\Delta^2)$,
Eq. (\ref{ii)}) is verified.

Besides, the polynomiality condition is formally fulfilled
by the GPD defined in Eq. (\ref{usu}), although the present
accuracy of the model, explicitly written in the latter equation,
does not allow to really check polynomiality, due to the already
mentioned effects of the Munich Symmetry \cite{mun}.

The definition of  $H_q(x,\xi,\Delta^2)$
in terms of CQM
wave functions can be generalized
to other GPDs, and the relation
of the latter quantities with other form factors (for example the magnetic
one) and
other PDFs (for example the polarized quark density)
can be recovered.
Therefore the proposed scheme allows one to calculate the
GPDs by using any CQM.

With respect to Eq. (\ref{usu}),
a few caveats are necessary.

i) One should keep in mind that
Eq. (\ref{usu}) is a NR result, holding for $\xi^2 \ll 1$, 
under the conditions $\vec k^2 \ll m^2$,
$\vec \Delta^2 \ll m^2$.
If one wants to treat more general processes,
the NR limit should be relaxed by 
taking into account relativistic corrections.
In this way,
at the same time, an expression to 
evaluate $E_q(x,\xi,\Delta^2)$ could be 
obtained. Since our main aim here is to
describe our approach, rather than to obtain realistic estimates,
we postpone to a later publication the discussion
of a relativistic model, which will permit to study
the full $\Delta^2$- and $\xi$-range, together
with the GPD $E_q(x,\xi,\Delta^2)$. 

ii) The Constituent Quarks are assumed to be point-like.

iii) If use is made of a CQM, containing only
constituent quarks (and also antiquarks in the case of mesons),
only the quark (and antiquark) contribution to the
GPDs can be evaluated, i.e.,
only the region $x \geq \xi$ 
(and also $x \leq -\xi$ for mesons) can be explored.
In order to introduce the study of the ERBL region
($ - \xi \leq x \leq \xi$), so that
observables like cross-sections, spin asymmetries and so on
can be calculated,
the model has to be enriched.

iv) In actual calculations, the evaluation of 
Eq. (\ref{usu}) requires the choice of a reference frame.
In the following, the Breit Frame will be chosen, where one has
$\Delta^2 = - \vec \Delta^2$ and, in the NR limit we are studying,
one finds $\sqrt{2} \bar P ^+ \rightarrow M$. It happens therefore that,
in the argument of the $\delta$ function
in Eq. (\ref{usu}), 
the $x$ variable for the valence quarks
is not defined in its natural support, i.e. it can be
larger than 1 and smaller than $\xi$.
Several prescriptions have been proposed in the past
to overcome such a difficulty in the standard PDFs case \cite{jaro,trvv}.
Although the
support violation is small for the calculations that will be shown here,
it has to be reported as a drawback of the approach.

The issue iii) will be discussed in the next sections,
by relaxing the condition ii) and allowing for a finite size
and composite structure of the constituent quark.

\section{GPDs in a Constituent Quark Scenario}

The procedure described in the previous section,
when applied in the standard
forward case, has been proven to be able to
reproduce the gross features of PDFs
\cite{trvv}.
In order to achieve a better agreement with data, the approach
has to be improved.

In a series of previous papers, it has been shown that
unpolarized
\cite{scopetta1} and polarized \cite{scopetta2} DIS data
are consistent with a low energy scenario
dominated by composite constituents
of the nucleon.
This was obtained
using a simple picture of the constituent quark as a
complex system of point-like partons, and thus constructing the 
forward parton distributions by
a convolution between constituent quark momentum distributions
and constituent quark structure functions.
The latter quantities were obtained by using updated phenomenological
information in a scenario firstly suggested by Altarelli,
Cabibbo, Maiani and Petronzio (ACMP) already in the seventies
\cite{acmp}. 

Following the same idea, in this section
a model for the reaction mechanism of an off-forward process,
such as DVCS, where GPDs could be measured,
will be proposed. 
As a result, a convolution formula giving
the proton $H_q$ GPD in terms of a constituent quark off-forward momentum 
distribution,
$H_{q_0}$,
and of a GPD of the constituent quark $q_0$ itself, 
$H_{q_0q}$,
will be derived.

It is assumed that the hard scattering with the virtual photon
takes place on a parton of a spin $1/2$ target, made of
complex constituents. This can be the case of a spin $1/2$ nucleus,
such as $^3He$, or of the proton, if this is assumed to be made
of composite constituent quarks.
The latter situation is the one we are interested in.

The scenario we are thinking of is depicted in Fig. 1 for the 
special case of DVCS, in the handbag
approximation. One parton ({\sl{current}} quark)
with momentum $k$,
belonging to a given constituent
of momentum p, interacts with the probe
and it is afterwards reabsorbed, with momentum
$k+\Delta$, by the same constituent,
without further re-scattering with the recoiling system
of momentum $P_R$.
We suggest here an analysis of the process 
which is quite similar to the usual IA approach
to DIS off nuclei \cite{muld,fs,cio}.

In the class of frames chosen in section 2,
and in addition to the kinematical variables,
$x$ and $\xi$, already defined,
one needs a few more to describe the process.
In particular,
$x'$ and $\xi'$, for the ``internal''
target, i.e., the constituent quark, have to be introduced.
The latter quantities can be obtained defining the ``+''
components of the momentum $k$ and $k + \Delta$ of the struck parton
before and after the interaction, with respect to
$\bar P^+$ and $\bar p^+ = {1 \over 2} (p + p')^+$:
\begin{eqnarray}
k^+ & = & (x + \xi ) \bar P^+ =  (x' + \xi') \bar p^+~,
\\
(k+\Delta)^+ & = & 
(x - \xi ) \bar P^+ =  (x' - \xi') \bar p^+~.
\end{eqnarray}
From the above expressions, $\xi'$ and $x'$ are immediately obtained as
\begin{eqnarray}
\xi' & = & - { \Delta^+ \over 2 \bar p^+}
\\
x' & = & {\xi' \over \xi} x
\end{eqnarray}
and, since $\xi = - \Delta^+ / (2 \bar P^+)$, 
if $\tilde z = p^+/P^+$, one also has
\begin{eqnarray}
\xi ' = {\xi \over \tilde z( 1 + \xi ) - \xi}~.
\label{xi1}
\end{eqnarray}

These expressions have been already found and used
in the IA analysis of DVCS off the deuteron
\cite{cano} and, in general, off nuclei \cite{kma}.

In order to derive a convolution formula,
a standard procedure will be adopted \cite{muld,fs,cio}.
In Eq. (\ref{int_1}),
two complete sets of states, corresponding
to the interacting constituent and to the recoiling system, are 
properly inserted 
to the left and right-hand sides of the quark operator:
\begin{eqnarray}
{H_q(x,\xi,\Delta^2)} & = &  
{\bra P'+ |} \,\,\sum_{\vec P_R',S_R',\vec p',s'}
{ \{ | P_R' S_R' \ket | p' s' \ket \}} 
\{ \bra P_R' S_R' |  {\bra p' s'|} \} 
\nonumber
\\
& &
%{
%\sum_i \sum_{r',r,\vec k}  
%\delta \left ( x + { \xi \over 2} - {k^+ \over \bar P^+} \right )
%u_{r'}^+(\vec k + \vec \Delta)
%b^+_{i,r'}(\vec k + \vec \Delta)
%b_{i,r}(\vec k) u_{r}(\vec k)}
\int 
{d \vec k \over (2 \pi )^3 }
{ k^+ \over k_0} 
\delta \left( {k^+ \over \bar P^+} - (x+\xi) \right)
{ 1 \over V}
\sum_\lambda 
 b_\lambda^\dag(\vec k + \vec{\Delta})
b_\lambda(\vec k )
\nonumber
\\
& &
\sum_{\vec P_R,S_R,\vec p,s}
\{ | P_R S_R \ket {| p s \ket \}} 
{\{ \bra P_R S_R |  \bra p s| \} 
\,\,\,\,|  P +  \ket}~,
\nonumber
\end{eqnarray}
and since, using IA,
\begin{eqnarray}
{\{ \bra P_R S_R |  \bra p s| \} 
| P S \ket} = {\bra P_R S_R, p s  
| P S \ket } (2 \pi)^3 \delta^3 (\vec P - \vec P_R - \vec p)
\delta_{S,S_R\,s}~,
\nonumber
\end{eqnarray}
a convolution formula,
valid for any GPD $H_q$ of the spin 1/2 complex target 
(the proton in the present case) in terms of the GPD $H_{q_0 q}$ of spin
1/2 structured constituents (the constituent quark, in the present case),
$q_0$,
is readily obtained:
\begin{eqnarray}
H_q(x,\xi,\Delta^2) & = & \sum_{q_0} \int dE \int d \vec p
\, P_{q_0}(\vec p, \vec p + \vec \Delta, E ) {\xi' \over \xi}
H_{q_0 q}(x',\xi',\Delta^2)~.
\label{spec}
\end{eqnarray}
In the above equation,
$E$ is the excitation energy of the recoiling system and
$P_{q_0} (\vec p, \vec p + \vec \Delta, E ) $
the one-body off-diagonal spectral function
for the constituent quark $q_0$ in the proton:
\begin{eqnarray}
P_{q_0}(\vec p, \vec p + \vec \Delta, E)  = 
%{1 \over 2} \sum_S 
\sum_{\vec S_R,s}
\bra \vec P'S | (\vec P - \vec p) S_R, (\vec p + \vec \Delta) s\ket 
\bra (\vec P - \vec p) S_R,  \vec p s| \vec P S \ket
\nonumber
% \\
% & &  
\, \delta(E - E_R)~.
\nonumber
\end{eqnarray}
If the $E$-dependence of $\xi'$, i.e.,
the $E$-dependence of $\tilde z$ (cf. Eq. (\ref{xi1}))
is disregarded in Eq. (\ref{spec}), so that
the one-body off-diagonal momentum distribution
\begin{equation}
n_{q_0}(\vec p, \vec p + \vec \Delta) =
\int dE P_{q_0}(\vec p, \vec p + \vec \Delta, E )~
\label{clo}
\end{equation}
is recovered, Eq.(\ref{spec}) can be written in the form
\begin{eqnarray}
H_{q}(x,\xi,\Delta^2) & \simeq & 
\sum_{q_0} \int d \vec p
\, n_{q_0}(\vec p, \vec p + \vec \Delta) {\xi' \over \xi}
H_{q_0 q}(x',\xi',\Delta^2) = \nonumber
\\
& = &  
\sum_{q_0} \int_x^1 {dz \over z} \int d \vec p
\, n_{q_0}(\vec p, \vec p + \vec \Delta) 
\delta \left( z - { \xi \over \xi' }  \right)
H_{q_0 q} \left( { x \over z }, { \xi \over z },\Delta^2 \right)~.
\label{pnk}
\end{eqnarray}
Taking into account that 
\begin{equation}
z - { \xi \over \xi'} = z - [ \tilde z ( 1 + \xi  ) - \xi ]
= z + \xi - { p^+ \over P^+} ( 1 + \xi )
= z + \xi  - { p^+ \over \bar P^+}~,
\end{equation}
Eq. (\ref{pnk}) can also be written in the form:
\begin{eqnarray}
H_{q}(x,\xi,\Delta^2) =  
\sum_{q_0} \int_x^1 { dz \over z}
H_{q_0}(z, \xi ,\Delta^2 ) 
H_{q_0 q} \left( {x \over z},
{\xi \over z},\Delta^2 \right)~,
\label{main}
\end{eqnarray}
where 
\begin{equation}
H_{q_0}(z, \xi ,\Delta^2 ) =  \int d \vec p
\, n_{q_0}(\vec p, \vec p + \vec \Delta) 
\delta \left( z + \xi  - { p^+ \over \bar P^+ } \right)
\label{hq0}
\end{equation}
is to be evaluated in a 
given CQM, according to Eq. (\ref{usu}),
for $q_0=u_0$
or $d_0$, while $H_{q_0 q}( {x \over z},
{\xi \over z},\Delta^2)$ is the constituent quark
GPD, which is still to be discussed and
will be modelled in the next section. One should notice that
the forward limit of Eq. (\ref{main}) gives the 
expression which is usually found,
for the parton distribution $q(x)$, in the IA analysis of
unpolarized DIS off nuclei \cite{muld,fs,cio}. 
In fact, if in Eq. (\ref{main})
the index $q_0$, labelling a constituent quark, is replaced 
by the index $N$, labelling a nucleon in the nucleus, in the forward limit
one finds the well known result:
\begin{eqnarray}
q(x) =  H_q(x,0,0) =
\sum_{N} \int_x^1 { dz \over z}
f_{N}(z) \,
q_{N}\left( {x \over z}\right)~,
\label{mainf}
\end{eqnarray}
where 
\begin{equation}
f_{N}(z) = H_{N}(z, 0 ,0 ) =  \int d \vec p
\, n_{N}(\vec p) 
\delta\left( z - { p^+ \over \bar P^+ } \right)
\label{hq0f}
\end{equation}
is the light-cone momentum distribution of the nucleon $N$
in the nucleus and $q_N(x)=
H_{N q}( x , 0, 0)$
is the distribution
of the quark of flavor $q$ 
in the nucleon $N$.

\section{A model for the GPDs of the Constituent Quark}

The crucial problem now
is the definition of $H_{q_0 q}( {x \over z},
{\xi \over z},\Delta^2)$, the constituent quark
GPD, appearing in Eq. (\ref{main}). 

As usual, we can start modelling this quantity
thinking first of all to its forward limit, 
where the constituent quark
parton distributions have to be recovered.
As we said in the previous section,
in a series of papers
\cite{scopetta1,scopetta2} 
a simple picture of the constituent quark as a
complex system of point-like partons
has been proposed, 
re-taking a scenario suggested by Altarelli,
Cabibbo, Maiani and Petronzio (ACMP)
\cite{acmp}.

Let us recall the main features of that idea.

The constituent quarks are themselves
composite objects whose structure functions are described by a set of functions
$\phi_{q_0q}(x)$ that specify the number of point-like partons 
of type $q$ which
are present in the constituent of type $q_0$, with fraction $x$ of its total
momentum. We will hereafter call these functions, generically, the structure
functions of the constituent quark.

The functions describing the nucleon parton distributions are expressed in
terms of the independent $\phi_{q_0q}(x)$ and of the constituent 
density distributions ($q_0=u_0,d_0$) as,
\bq
q(x,Q^2)=\sum_{q_0}\int_x^1 {dz\over z} 
q_0(z,Q^2) \phi_{q_0q} \left({x \over z},Q^2 \right)~,
\label{fx}
\eq
where $q$ labels the various partons, i.e., valence quarks  
($ u_v,d_v$), sea quarks ($u_s,d_s, s$), sea antiquarks ($\bar u,\bar d, \
\bar
s$) and gluons $g$.

The different types and functional forms of the structure functions of the 
constituent quarks are derived from three very natural assumptions 
\cite{acmp}:
\begin{itemize} 
\item[  i)]The point-like partons are $QCD$ degrees of freedom, 
i.e. quarks, 
antiquarks and gluons;
\item[ ii)] Regge behavior for $x\rightarrow 0$ and duality ideas;
\item[iii)] invariance under charge conjugation and isospin.
\end{itemize}

These considerations define in the case of the valence quarks the following
structure function

\bq
\phi_{qq_v}({x},Q^2)
= { \Gamma(A + {1 \over 2}) \over 
\Gamma({1 \over 2}) \Gamma(A) }
{ (1-x)^{A-1} \over \sqrt{x} }.
\label{csf1}\eq
For the sea quarks the corresponding structure function becomes,

\bq
\phi_{qq_s}({x},Q^2)
= { C \over x } (1-x)^{D-1},\label{csf2}
\eq
and, in the case of the gluons, it is taken

\bq
\phi_{qg}({x},Q^2)
= { G \over x } (1-x)^{B-1}~.\label{csf3}
\eq

The last assumption 
of the approach relates to the scale at which the constituent quark 
structure is defined. We choose for it the so called hadronic
scale $\mu_0^2$ \cite{trvv,grv}. This hypothesis fixes $all$ 
the parameters of
the approach (Eqs. (\ref{csf1}) through (\ref{csf3})). 
The constants $A$, $B$, $G$
and the ratio $C/D$ are determined by the amount of momentum carried by the 
different partons, 
%We choose, 53.5 $\%$ by the valence quarks and 
%35.7 $\%$ by the gluons,
%which 
corresponding to a hadronic scale of $\mu_0^2=0.34$ GeV$^2$,
according to  
the parametrization of \cite{grv}. $C$ (or $D$) is fixed according to 
the value of $F_2$ at $x=0.01$ \cite{acmp},
and its value is chosen again according to \cite{grv}.
We stress that all these inputs are forced only by the updated
phenomenology, through the 2$^{nd}$ moments of PDFs. 
The values of the parameters obtained 
are listed in \cite{scopetta1}.

We, here, note that the unpolarized
structure function $F_2$ is rather insensitive to the change 
of the sea ($C$, $D$) and gluon ($B$, $G$) parameters.

The other ingredients appearing in Eq. ({\ref{fx}}), i.e., 
the density distributions for 
each constituent quark, are defined according to 
Eq. (\ref{usu}).

Now we have to generalize this scenario to describe off-forward 
phenomena. Of course, the forward limit of our
GPDs formula, Eq. (\ref{main}), has to be given by 
Eq. ({\ref{fx}}).
By taking the forward limit of Eq. (\ref{main}), one obtains:

\begin{eqnarray}
H_{q}(x,0,0) & = & 
\sum_{q_0} \int_x^1 { dz \over z}
H_{q_0}(z, 0, 0 ) 
H_{q_0 q} \left( {x \over z},0,0 \right) 
\nonumber \\
& = & 
\sum_{q_0} \int_x^1 { dz \over z}
q_0(z) 
H_{q_0 q} \left( {x \over z},0,0 \right)~,
\label{for}
\end{eqnarray}

so that, in order for the latter to coincide with Eq. 
(\ref{fx}), one must have
$H_{q_0 q}( x,0,0) \equiv  
\phi_{q_0q}(x)$.

In such a way, through the ACMP prescription, the forward
limit of the unknown constituent quark GPD
$H_{q_0 q}( {x \over z},
{\xi \over z},\Delta^2)$ can be fixed.

Now the off-forward behavior of the Constituent Quark GPDs 
has to be modelled.

This can be done in a natural way by using the ``$\alpha$-Double
Distributions'' (DD's) language proposed by Radyushkin \cite{rag,radd}.
DD's, $\Phi(\tilde x, \alpha, \Delta^2)$,
are a representation of GPDs which automatically guarantees
the polynomiality property.
GPDs can be obtained from DD's after a proper integration.
%Such a relation is shown diagramatically in Fig. in the handbag
%approximation.
In constructing models, the DD's can represent a more appropriate
language with respect to GPDs. In fact, the hybrid
character of GPDs, which are something
in between parton densities $q(x)$ and
distribution amplitudes $\phi(\alpha)$, is naturally emphasized 
when the latter are obtained from DD's.
The DD's do not depend on the skewedness parameter $\xi$; rather,
they describe how the {\sl{total}}, $P$, and {\sl{transfer}}, $\Delta$,
momenta are shared between the interacting and final partons, by means
of the variables $\tilde x$ and $\alpha$, respectively.
 As it can be argued
from Fig. 2, 
where
the DDs representation of GPDs is illustrated schematically, 
and as it is explained in \cite{rag,radd},
parton densities are recovered in the forward, $\Delta=0$ limit,
while meson distribution amplitudes are obtained in the $P=0$
limit of DD's. In some cases, such a transparent physical interpretation,
together with the symmetry properties which are typical
of distribution amplitudes ($\alpha \rightarrow - \alpha$ symmetry),
allows a direct modelling, already developed in \cite{rad1}.

The relation between any GPD $H$, defined {\sl \`a la Ji},
for example the one we need, i.e. 
$H_{q_0q}$ for the constituent quark target,
is related to the $\alpha$-DD's, which we call
$\tilde \Phi_{q_0 q} (\tilde x, \alpha,\Delta^2)$ for the constituent
quark,
in the following way
\cite{rag,radd}:
\begin{eqnarray}
H_{q_0q}(x,\xi,\Delta^2) = \int_{-1}^1 d\tilde x
\int_{-1 + |\tilde x|}^{1-|\tilde x|} 
\delta(\tilde x + \xi  \alpha - x)
\tilde \Phi_{q_0 q} (\tilde x, \alpha,\Delta^2) d \alpha~.
\label{hdd}
\end{eqnarray}

With some care, the expression above can be integrated over 
$\tilde x$ and the result is explicitly given in 
\cite{rag}.
The DDs fulfill the relation:
\begin{equation}
\tilde \Phi_{q_0q} (\tilde x, \alpha,\Delta^2) =
\tilde \Phi_{q_0q} (\tilde x, -\alpha,\Delta^2)~,
\end{equation}
and the polynomiality condition \cite{jig}.

In \cite{radd}, a factorized
ansatz is suggested for the DD's: 
\begin{equation}
\tilde \Phi_{q_0 q} (\tilde x, \alpha,\Delta^2) =
h_{q} (\tilde x, \alpha,\Delta^2)
\Phi_{q_0 q} (\tilde x) 
F_{q_0}(\Delta^2)~,
\label{ans}
\end{equation}
with 
the $\alpha$ dependent
term, 
$h_{q} (\tilde x, \alpha,\Delta^2)$, 
which has the character of a mesonic amplitude,
fulfilling the relation:
\begin{equation}
\int_{-1 + |\tilde x|}^{1-|\tilde x|} h_{q} 
(\tilde x, \alpha,\Delta^2) d \alpha = 1~.
\label{hnor}
\end{equation}
Besides, in Eq. (\ref{ans}) 
$
\Phi_{q_0 q} (\tilde x) 
$
represents the forward density
and, eventually, 
$
F_{q_0}(\Delta^2)
$
the constituent quark form factor.

One immediately realizes that the GPD of the constituent quark,
Eq. (\ref{hdd}), with the factorized form Eq. 
(\ref{ans}) and the normalization Eq. (\ref{hnor}), fulfills the
crucial constraints of GPDs, i.e., the forward limit,
the first-moment and the polynomiality condition,
the latter being automatically verified in the DD's
description. 

In the following, we will assume for the constituent quark GPD the 
above factorized form, so that we need to model the three functions
appearing in Eq. (\ref{ans}), according to the description
of the reaction mechanism we have in mind.

For the amplitude $h_q$, use will be made of
one of the simple normalized forms suggested in
\cite{radd}, 
on the bases of the symmetry
properties of DD's:
\begin{equation}
h_{q}^{(1)}(\tilde x, \alpha) = {3 \over 4} { (1 - \tilde x)^2 - \alpha^2 
\over (1 - \tilde x)^3 }~.
\label{h}
\end{equation}

Besides,
since
we will identify
quarks for $x \ge \xi/2$, pairs for $ x \le |\xi/2|$,
antiquarks for  $x \le -\xi/2$, and, since
in our approach the forward densities
$\Phi_{q_0q} (\tilde x)$ have to be given by
the standard $\Phi$ functions of the 
$ACMP$ approach, Eqs. (\ref{csf1})--(\ref{csf3}),
one has, for the DD of flavor $q$
of the constituent quark:
\begin{eqnarray}\label{sr}
\tilde \Phi_{q_0 q} (\tilde x, \alpha,\Delta^2) =
\cases{
(h_{q}(\tilde x,\alpha) \Phi_{q_0q_v}(\tilde x)
+
h_{q}(\tilde x,\alpha) \Phi_{q_0q_s}(\tilde x)) F_{q_0}(\Delta^2) 
 & 
for $\tilde x \ge 0$ \cr
- h_{q}(- \tilde x,\alpha) \Phi_{q_0q_s}(- \tilde x)
F_{q_0}(\Delta^2) 
 & for $\tilde x < 0$ \cr}
\label{dd}
\end{eqnarray}

The above definition, due to Eq. (\ref{hnor}), 
when integrated over $\alpha$
gives the correct limits \cite{rag}:
\begin{equation}
\int_{-1 + \tilde x}^{1-\tilde x} \tilde \Phi_{q_0q} 
(\tilde x, \alpha,\Delta^2=0)|_{\tilde x > 0} 
d \alpha = \Phi_{q_0q} (\tilde x)~,
\end{equation}
and
\begin{equation}
\int_{-1 + |\tilde x|}^{1-|\tilde x|} \tilde \Phi_{q_0q} 
(\tilde x, \alpha,\Delta^2=0)|_{\tilde x < 0} d \alpha = - \Phi_{q_0 \bar q} 
(-\tilde x)~.
\end{equation}

Eventually, as a f.f. we will take a monopole form
corresponding to a constituent quark size $r_Q \simeq 0.3 fm$:

\begin{equation}
F_{q_0}(\Delta^2) = { 1 \over 1 - {\Delta^2 r_Q^2
\over 6 }}~,
\label{ffacmp}
\end{equation}
a scenario strongly supported by the analysis of \cite{psr}.

By using such a f.f. and Eq. (\ref{h}), together with the standard
ACMP $\Phi$'s, Eqs. (\ref{csf1}) and
(\ref{csf2}), in Eq. (\ref{dd}), and inserting the obtained
$\Phi_{q_0 q} (\tilde x, \alpha,\Delta^2)$ 
into Eq. (\ref{hdd}), 
the constituent quark GPD in the ACMP scenario can be eventually calculated.

\section{Results and discussion}

In this section we present the results obtained 
for the GPD $H_q$ of the proton,
for $\xi^2 \ll 1$ and $\vec \Delta^2 \ll m^2$, 
according to the approach described so far.
The main equation to be evaluated is 
Eq. (\ref{main}), written again here below for the sake of clarity:
\begin{eqnarray}
H_{q}(x,\xi,\Delta^2) =  
\sum_{q_0} \int_x^1 { dz \over z}
H_{q_0}(z, \xi ,\Delta^2 ) 
H_{q_0 q}\left( {x \over z},
{\xi \over z},\Delta^2 \right)~.
\nonumber
\end{eqnarray}

In the above equation,
the quantity $H_{q_0q}$, the constituent quark GPD, is modelled
according to the arguments described in the previous section.
This means that it is obtained evaluating Eq. (\ref{hdd}),
where the DD of the constituent quark,
$\tilde \Phi_{q_0 q} (\tilde x, \alpha,\Delta^2)$ ,
is given by Eq. (\ref{dd}), calculated in turn through
the f.f., Eq. (\ref{ffacmp}), the function
$h_q$,  Eq. (\ref{h}), together with the standard
ACMP $\Phi$'s, Eqs. (\ref{csf1}) and
(\ref{csf2}).

The other ingredient in Eq. (\ref{main}), 
$H_{q_0}$, has been evaluated according to 
Eq. (\ref{hq0}):
\begin{eqnarray}
H_{q_0}(z, \xi ,\Delta^2 ) =  \int d \vec p
\, 
n_{q_0}(\vec p, \vec p + \vec \Delta) 
\delta \left( z + \xi  - { p^+ \over \bar P^+ } \right)~.
\nonumber
\end{eqnarray}
The calculation has been performed in the Breit frame,
where one has, in the NR limit studied, $\sqrt{2} \bar P^+ \rightarrow M$.
The off-diagonal momentum distribution appearing in the
formula above,
$n_{q_0}(\vec p, \vec p + \vec \Delta)$, defined in
Eq. (\ref{clo}), has been evaluated within 
the Isgur and Karl (IK) model \cite{ik}. 
The calculation is described in \cite{epj} and the main results
are listed again here for the reader's convenience.

In the IK CQM \cite{ik},
including contributions up to the $2 \hbar \omega$ shell, 
the proton state is given by the
following admixture of states
\begin{eqnarray}
|N \rangle = 
a_{\cal S} | ^2 S_{1/2} \rangle_S +
a_{\cal S'} | ^2 S'_{1/2} \rangle_{S} +
a_{\cal M} | ^2 S_{1/2} \rangle_M +
a_{\cal D} | ^4 D_{1/2} \rangle_M~,
\label{ikwf}
\end{eqnarray}
where the spectroscopic notation $|^{2S+1}X_J \rangle_t$, 
with $t=A,M,S$ being the symmetry type, has been used.
The coefficients were determined by spectroscopic properties to be
\cite{mmg}: 
$a_{\cal S} = 0.931$, 
$a_{\cal S'} = -0.274$,
$a_{\cal M} = -0.233$, $a_{\cal D} = -0.067$.
%The different components appearing in the momentum
%space wave functions, obtained from Eq. (\ref{ikwf})
%in the IK model, can be found in 
%\cite{mmg,ik2};
%for the h.o. model,
%the corresponding wave function
%in momentum space reduces to \cite{traini,mmg}
%\begin{eqnarray}
%\label{howf}
%\psi (\vec{k_{\rho}}, \vec{k_{\lambda}} ) 
%= { e^{- { { k_{\rho}^2 + k_{\lambda}^2} \over 2 \alpha^2 } }  
%\over \pi^{3/2} \alpha^3 }~,
%\end{eqnarray}
%where the h.o. parameter can be fixed
%to $\alpha^2=1.35 f^{-2}$
%in order to reproduce the low $t$ behavior of the charge
%FF, i.e., the r.m.s. value of the proton radius.
\\
The results for the 
GPD $H(x,\xi,\Delta^2)$, 
neglecting in (\ref{ikwf})
the small $D$-wave contribution, have been found to be
\cite{epj}:

\begin{eqnarray}
H_u(x,\xi,\Delta^2) 
& = & 3 { M \over \alpha^3}  
\left( {3 \over 
2 \pi } \right)^{3/2} 
e^{ { \Delta^2 \over 3 \alpha^2 }} 
\int dk_x \int dk_y \,
f_0(k_x,k_y,x,\xi,\Delta^2) \nonumber
\\
& \times &
\left [ f_s(k_x,k_y,x,\xi,\Delta^2)
+ {\tilde f} (k_x,k_y,x,\xi,\Delta^2) \right ]~,
\label{iku}
\end{eqnarray}

\begin{eqnarray}
H_d(x,\xi,\Delta^2) 
& = & 3 { M \over \alpha^3}  
\left( {3 \over 
2 \pi } \right)^{3/2} 
e^{{ \Delta^2 \over 3 \alpha^2 }} 
\int dk_x \int dk_y \,
f_0(k_x,k_y,x,\xi,\Delta^2) \nonumber
\\
& \times &
\left [ {1 \over 2} f_s(k_x,k_y,x,\xi,\Delta^2)
- {\tilde f} (k_x,k_y,x,\xi,\Delta^2)\right ]~,
\label{ikd}
\end{eqnarray}

for the flavors $u$ and $d$, respectively, 
with

\begin{eqnarray}
f_0(k_x,k_y,x,\xi,\Delta^2) =
{ \bar k_0 \over \bar k_0 + \bar k_z }
f_\alpha(\Delta_x,k_x)
f_\alpha(\Delta_y,k_y) f_\alpha(\Delta_z,\bar k_z)~,
\end{eqnarray}
\begin{eqnarray}
f_\alpha(\Delta_i,k_i) = e^{ - {1 \over \alpha^2} \left( {3 \over 2} k_i^2  
+ k_i\Delta_i \right) }~~,
\end{eqnarray}
\begin{eqnarray}
\bar{k_z} = { M^2(x + \xi)^2 - ( m^2 + k_x^2 + k_y^2)
\over 2 M ( x + \xi) }~,
\end{eqnarray}

\begin{eqnarray}
\label{fs}
f_s(k_x,k_y,x,\xi,\Delta^2) & = & {2 \over 3} a_s^2 + a_{s'}^2
\left[ {5 \over 6} - {k^2 \over \alpha^2} + { 1 \over 2} 
{k^4 \over \alpha^4} + { 2 \over {3 \alpha^2}}
\left ( - {\Delta^2 \over 3} + \vec \Delta \cdot \vec k \right )
\left ( {k^2 \over \alpha^2} -1 \right ) \right ] \nonumber
\\  
& + & a_M^2 \left [ {5 \over 12} - {1 \over 2} {k^2 \over \alpha^2}  
+ { 1 \over 4} {k^4 \over \alpha^4} +
{ 2 \over 9} {k \over \ \alpha^2} \sqrt{ { 9 \over 4} k^2 -
\Delta^2 + 3 \vec \Delta \cdot \vec k } \right .
\nonumber
\\
& + & \left .
{ 1 \over {3 \alpha^2}}
\left ( -{\Delta^2 \over 3} + \vec \Delta \cdot \vec k \right )
\left ( {k^2 \over \alpha^2} -1 \right ) \right ]
\\
& + & a_S a_{S'} { 2 \over \sqrt{3}} 
\left [ \left ( 1 - {k^2 \over \alpha^2} \right )
- { 2 \over {3 \alpha^2}}
\left ( -{\Delta^2 \over 3} + \vec \Delta \cdot \vec k \right )
\right ] \nonumber
\end{eqnarray}

\begin{eqnarray}
\label{ft}
\tilde f(k_x,k_y,x,\xi,\Delta^2) & = & - a_S a_{S'} { 2 \over \sqrt{3}} 
\left [ \left ( 1 - {k^2 \over \alpha^2} \right )
- { 2 \over {3 \alpha^2}}
\left ( - {\Delta^2 \over 3} + \vec \Delta \cdot \vec k \right )
\right ] \nonumber \\
& - & a_M a_{S'} { 1 \over \sqrt{2}} 
\left [ { 1 \over 6} - {k^2 \over \alpha^2} +
{1 \over 2} {k^4 \over \alpha^4}
- { 2 \over {3 \alpha^2}}
\left ( -{\Delta^2 \over 3} - \vec \Delta \cdot \vec k \right )
\right .
\nonumber
\\
& + & \left . { 2 k^2 \over {3 \alpha^4}}
\left ( -{\Delta^2 \over 3} + \vec \Delta \cdot \vec k \right )
\right ]
\end{eqnarray}
and $\bar k_0 = \sqrt{ m^2+k_x^2+k_y^2+\bar k_z^2}$, being $m \simeq
M/3$ the constituent quark mass.
Here we have used the notation $k^2=\vec k^2, \vec k = (k_x,k_y,k_z)$.

The harmonic oscillator parameter,
$\alpha$, of the IK model,
can be chosen so that the experimental
r.m.s. of the proton is reproduced by the slope,
at $\Delta^2=0$,
of the charge form factor.
Such a choice is performed as follows.
Integrating Eq. (\ref{main}) over $x$, one gets:
\bq
F_q(\Delta^2) = \sum_{q_0} 
F_{q_0}(\Delta^2)
F_{q_0q}(\Delta^2)~,
\label{ffc}
\eq
where 
$F_q(\Delta^2)$ is the
contribution, 
of the {\sl current} quark of flavor $q$,
to the
proton f.f.;
$F_{q_0}(\Delta^2)$ is the contribution,
of the 
{\sl point-like} constituent quark of flavor $q_0$,  
to the proton f.f.;
$F_{q_0q}(\Delta^2)$ is the contribution,
of the current quark of
flavor $q$,
to the f.f. of the {\sl composite} constituent quark of 
flavor $q_0$.

The latter is given by Eq. (\ref{ffacmp}),
while in the IK model one has \cite{mmg}:
\bq
F_{q_0}(\Delta^2)= 
\left ( 1 - {a \over \alpha} \Delta^2
- {b \over \alpha^2} \Delta^4 \right )
e^{\Delta^2 \over 6 \alpha^2}~,
\label{ffik}
\eq
with $a\simeq 0.015$ and $b\simeq 0.0001$.
Imposing that the slope of the f.f., Eq.(\ref{ffc}),
reproduces at $\Delta^2=0$ the experimental
proton r.m.s., a value of $\alpha=1.18 f^{-1}$ is obtained.
Such a form factor reproduces well the data at the low values of
$\Delta^2$ which are accessible in the present approach.
For higher values of $\Delta^2$, it would not be realistic
\cite{mmg}.

Results for the $u$-quark $H$ distribution, at the scale of the model
$\mu_o^2$,
are shown in Figs 3 to 5.
In Fig. 3, it is shown
for $\Delta^2=-0.1$ GeV$^2$ and $\xi=0.1$.
One should remember that the present approach does not allow
to estimate realistically the region $-\Delta^2 \ge m^2 \simeq 0.1$ GeV$^2$,
so that we are showing here the result corresponding
to the highest possible $\Delta^2$ value.
Accordingly, the maximum value of 
the skewedness 
is therefore
$\xi \simeq 0.17$ (cf. Eq. (\ref{xim})), fulfilling
the requirement $\xi^2 \ll 1$.
The dashed curve represents what is obtained in the pure
Isgur and Karl model, i.e., by evaluating Eq. 
(\ref{iku}). One should notice that such a result, obtained
in a pure valence CQM, should 
vanish for $x \le \xi$ and for $x \ge 1$. The small tails
which are found in these forbidden regions represent the amount
of support violation of the approach.
In particular, for the shown values of $\Delta^2$ and $\xi$, a violation
of 2 $\%$ is found. In general, in the accessible region the violation
is never larger than few percents.
The full curve in Fig. 3 represents the complete result of the present
approach, i.e., the evaluation of Eq. (\ref{main}) following the steps
and using the ingredients described in this section and in the 
previous one. 
A relevant 
contribution is found to lie in the ERBL region,
in agreement with other estimates \cite{pog}.
As already explained, the knowledge
of GPDs in the ERBL region is a crucial prerequisite for
the calculation of all the cross-sections and the observables
measured in the processes where GPDs contribute.  
We notice that the ERBL region is accessed here, with
respect to the approach of Ref. \cite{epj}
which gives the dashed curve, thanks
to the constituent structure which has been introduced
by the ACMP procedure.

In Fig. 4, special emphasis is devoted to show the
$\xi$-dependence of the results. For the allowed
$\xi$ values, $H_u(x,\xi,\Delta^2)$,
evaluated using our main formula, Eq. (\ref{main}),
is shown for
four different values of $x$. It is clearly seen that such
a dependence is strong in the ERBL region,
while it is rather mild in the DGLAP region, in good agreement
with other estimates \cite{pog}.
To allow for a complete view of the outcome of our approach,
in Fig. 5 the $x$ and $\xi$ dependences are shown together
in a 3-dimensional plot.

The results shown so far are associated with
the low scale of the model, the hadronic scale $\mu_0^2$,
fixed to the value 0.34 GeV$^2$, as discussed in Section 4.
As an illustration, in Fig. 6 and 7 the Next to Leading Order QCD-evolution 
of the Non Singlet (NS), valence $u$ distribution, up
to a scale of $Q^2=10$ GeV$^2$, is shown.
One should notice that any NS distribution is symmetric in $x$,
due to its definition and in agreement with the conventions used:
\be
H_q^{NS}(x,\xi,\Delta^2) & = & H_q(x,\xi,\Delta^2) -
H_{\bar q}(x,\xi,\Delta^2) =
\nonumber
\\
& = & H_q(x,\xi,\Delta^2) + H_q(-x,\xi,\Delta^2)~.
\ee
In evolving the model results,
the approach of Ref. \cite{evo} has been applied
and a code kindly provided by A. Freund has been used.
The evolution clearly shows a strong enhancement of the ERBL region. 

We have therefore developed a scheme which provides us
with the GPD $H$ in the full $x$ range. 
This is obtained thanks to the constituent quark
structure, implemented dressing the three quarks of a CQM, 
where initially only the DGLAP region of GPDs was accessible.
This is an important development with respect
to previous work, a prerequisite for
any attempt to calculate cross sections and asymmetries
of related processes.
The next step of our studies will be indeed to use the obtained
GPDs for the evaluation of cross sections
which have been recently measured
\cite{hermes,clas} or will be measured soon.
The estimate of cross sections is presently in progress,
together with that of relativistic corrections,
which will permit to enlarge the kinematical range 
(basically the $\Delta^2$ and the related $\xi$-ranges) where
our results can be applied to predict or to describe the data.
For the time being, the comparison with data of our estimates
is therefore not possible. 

Anyway, as already said, the present approach fulfills several
theoretical constraints.
First of all, the forward limit, Eq. (\ref{fx}), provides us with
a reasonable description of quark densities
(see Ref. \cite{scopetta1}, where the IK model together with the
ACMP mechanism has been
applied to unpolarized DIS
\footnote{In that work \cite{scopetta1}, we showed
that more sophisticated quark models were producing an excellent description
of the data \cite{iac}.}).
Secondly, the $x-$integral of our GPD $H$ turns out to be, formally
and numerically, $\xi$ independent, satisfying therefore 
the polynomiality condition for the first moment, and providing us with
a proton form factor, Eq. (\ref{ffik}), in good agreement
with the data in the low-$\Delta^2$ region which is studied here. 
The main theoretical drawback is the already discussed support-violation,
which in any case affects 
our findings,
in the kinematical region under investigation,
by a few percents at most.
Another theoretical constraint which is satisfied by the present approach
is the inequality:
\bq
H_q(x,\xi,\Delta^2)\le
\sqrt{ q(x_1)q(x_2) / (1 - \xi^2)}~,
\eq
where $x_1 = (x + \xi)/(1 + \xi)$ and
$x_2 = (x - \xi)/(1 - \xi)$, proven in \cite{radd}.
As an illustration,
the validity of the above inequality
is shown in Fig. 8 for the $u$-flavor, and for 
$\Delta^2 = -0.1$ GeV$^2$ and $\xi=0.1$.

Other theoretical constraints, such as the Ji sum rule
\cite{ji1}, require the knowledge of the other unpolarized GPD, 
$E$, which has not yet
been calculated in the present scheme.
$E$ arises naturally in a relativistic framework, where 
we plan to calculate it. In fact, a relativistic calculation will
permit us to overcome the support problem and, at the same time, 
to enlarge the kinematical range of our predictions.
Once a relativistic CQM is used to predict GPDs in the DGLAP
region, the structure of the constituent quark can be introduced
to access the ERBL region, so that 
the Ji sum rule and other theoretical
constraints can be checked. 
Work is being carried out in that direction.

\section{Conclusions}

Quark models have been extremely useful for understanding many features of
hadrons, even in the DIS regime. 
In previous work, a thorough analysis of both polarized
and unpolarized data have shown that constituent quarks cannot be considered
elementary when studied with high-energy probes. The first feature
found was the need for evolution, i.e. the constituent quarks at the
hadronic scale have to be undressed by incorporating bremsstrahlung
in order to reach the Bjorken regime, but this was not all.
A second feature, we found, was that the constituent quarks
should be endowed with soft structure in order to approach the data.
Thus, the constituent quarks appear, when under scrutiny by high-energy
probes, as complex systems, with a very different 
behavior from the current quarks of the basic theory.
These features, which we found in structure functions, have also
been recently discussed in form factors \cite{psr}.

The aim of the present work has been to generalize the formalism of composite
constituent quarks to study the generalized parton distributions.
We here develop a formalism which expresses the hadronic GPDs 
in terms of constituent quarks GPDs by means of appropriate
convolutions. In order to be able 
to predict experimental results we have defined
a model which incorporates phenomenological features
of various kinematical regimes. The model is based on Radyushkin's
factorization ansatz, thus our constituent quark GPDs are defined in terms
of the product of three functions: i) the constituent quark structure 
function, where we use the ACMP proposal \cite{acmp}; ii)
Radyushkin's double distributions \cite{radd}; iii) 
constituent quark form factor as suggested in Ref. \cite{psr}.
Once these GPDs are defined in this way, we have developed the scheme
to incorporate them into any nucleon model
by appropriate convolution. In order to show the type of predictions
to which
our proposal leads, we have used here, as an illustration,
a naive model of hadron structure,
namely the IK \cite{ik} model. However, in this latest step
of our scheme, any
non-relativistic (or relativized) model can be used to define the hadronic
GPDs.

Looking at our results, we found that the present scheme transforms 
a hadronic model, in whose original description only
valence quarks appear, into one containing
all kinds of partons (i.e., quarks, antiquarks and gluons).
Moreover, the starting model produces no structure in
the ERBL region, while after the structure of the constituent quark 
has been incorporated, it does. The completeness
of the $x$-range,
for the allowed
$\Delta^2$ and $\xi$, of the present description,
is a prerequisite for
the calculation of cross-sections and other observables in a wide
kinematical range.
To this respect we recall that
our findings hold for $\Delta^2 << m^2$ and 
$\xi^2 \ll 1$. Nevertheless, relativistic corrections,
which permit to access a wider kinematical region, could be included
in the approach. Our aim here has been mainly the illustration
of our scheme, and the inclusion of relativistic corrections, together
with the use of more sophisticated models, has been postponed and
will be shown elsewhere.

We have reminded the reader
of the calculation for the diagonal structure functions and
form factors to see how in these cases, where experimental data are available,
our scheme leads, even with a naive quark model, 
to a reasonable description of 
the data.
Thereafter, we have proceeded to calculate the GPDs of physical interest
to guide the preparation and analysis of future experiments.

This work is the continuation of an effort to construct a scheme 
which describes the properties of hadrons
in different kinematical and dynamical scenarios. Our description can
never be a substitute of Quantum Chromodynamics, but, before a solution
of it can be found, it may serve to guide experimenters to
physical processes where the theory might show interesting features, worthy
of a more fundamental effort.

The approach here presented can be extended to light nuclei and fragmentation 
functions, and work is being carried out also in these directions.

\acknowledgements

We acknowledge useful correspondence with
A. Radyushkin and helpful discussions with G. Salm\`e.
S.S. thanks the Department of Theoretical Physics of the Valencia
University, where part of this work has been done, for a warm hospitality
and financial support. V.V. thanks the Department of Physics 
of the University of Perugia for hospitality and support, and L. 
Kaptari for his help and interesting discussions.
S.S. thanks the Department of Energy's Institute for Nuclear Theory
of the University of Washington, Seattle, WA, for its hospitality
during the Program ''GPDs and hard exclusive processes'',
and the Department of Energy for partial support during
the completion of this work.

\newpage
\appendixonfalse
\section*{Figure Captions}

\vspace{1em}\noindent
{\bf Fig. 1}:
The handbag contribution to the DVCS process
in the present approach.

\vspace{1em}\noindent
{\bf Fig. 2}:
Illustration of the 
relation between $\alpha$-DDs (a)
and GPDs (b).

\vspace{1em}\noindent
{\bf Fig. 3}:
The GPD H for the flavor $u$,
for $\Delta^2=-0.1$ GeV$^2$ and $\xi=0.1$,
at the momentum scale of the
model.
Dashed curve: result in the pure
Isgur and Karl model, Eq. 
(\ref{iku}). 
The small tails
which are found in the forbidden regions,
$x \le \xi$ and  $x \ge 1$,
represent the amount
of support violation of the approach.
Full curve: the complete result of the present
approach, Eq. (\ref{main}).

\vspace{1em}\noindent
{\bf Fig. 4}:
For the $\xi$ values
which are allowed 
at $\Delta^2 = -0.1$ GeV$^2$,
$H_u(x,\xi,\Delta^2)$,
evaluated using our main equation, Eq. (\ref{main}),
is shown for
four different values of $x$,
at the momentum scale of the model.
From top to bottom,
the dash-dotted line represents the GPD at $x=0.05$,
the full line at $x=0.1$, the dashed line at $x=0.2$,
and the long-dashed line at $x=0.4$. 

\vspace{1em}\noindent
{\bf Fig. 5}:
The $x$ and $\xi$ dependences of
$H_u(x,\xi,\Delta^2)$, for $\Delta^2 = -0.1$ GeV$^2$,
at the momentum scale of the model.

\vspace{1em}\noindent
{\bf Fig. 6}:
The Non Singlet (NS) $H$ GPD for valence $u$-quarks
at $\xi=0.1$ and $\Delta^2 = -0.1$ GeV$^2$, at the momentum scale of the
model, $\mu_0^2=0.34$ GeV$^2$ (dashed), and after NLO-QCD evolution
up to $Q^2=10$ GeV$^2$ (full).

\vspace{1em}\noindent
{\bf Fig. 7}:
The Non Singlet $H$ GPD for valence $u$-quarks
at $\xi=0.1$ and $\Delta^2 = -0.1$ GeV$^2$, evolved at NLO 
from the momentum scale of the
model, $\mu_0^2=0.34$ GeV$^2$, up to $Q^2=10$ GeV$^2$.

\vspace{1em}\noindent
{\bf Fig. 8}:
Full curve: the GPD
$H_u(x,\xi,\Delta^2)$, for $\Delta^2 = -0.1$ GeV$^2$
and $\xi = 0.1$,
at the momentum scale of the model; dashed curve:
the quantity $\sqrt{ u(x_1)u(x_2) / (1 - \xi^2)}$,
where $x_1 = (x + \xi)/(1 + \xi)$ and
$x_2 = (x - \xi)/(1 - \xi)$ (see text).

\newpage

\begin{figure}[h]
\vspace{6.6cm}
\includegraphics{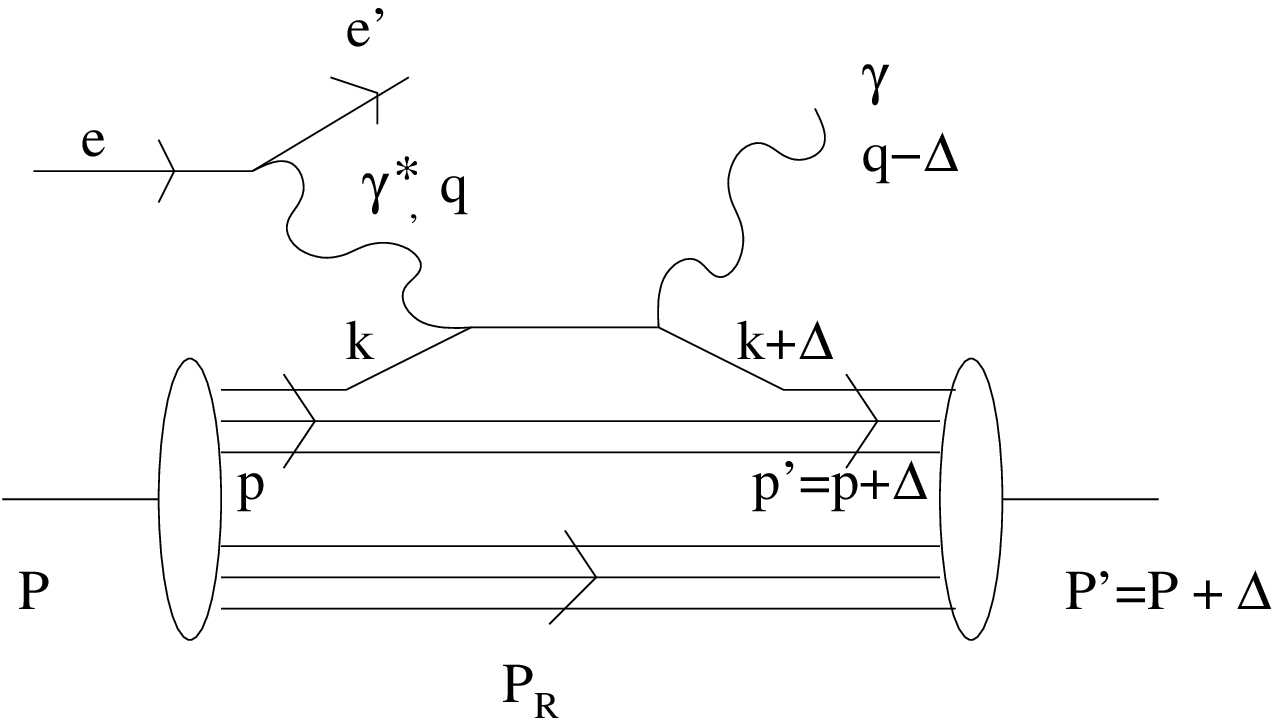}
\caption{}
\end{figure}

\vskip 2cm

\begin{figure}[h]
\vspace{6.6cm}
\includegraphics{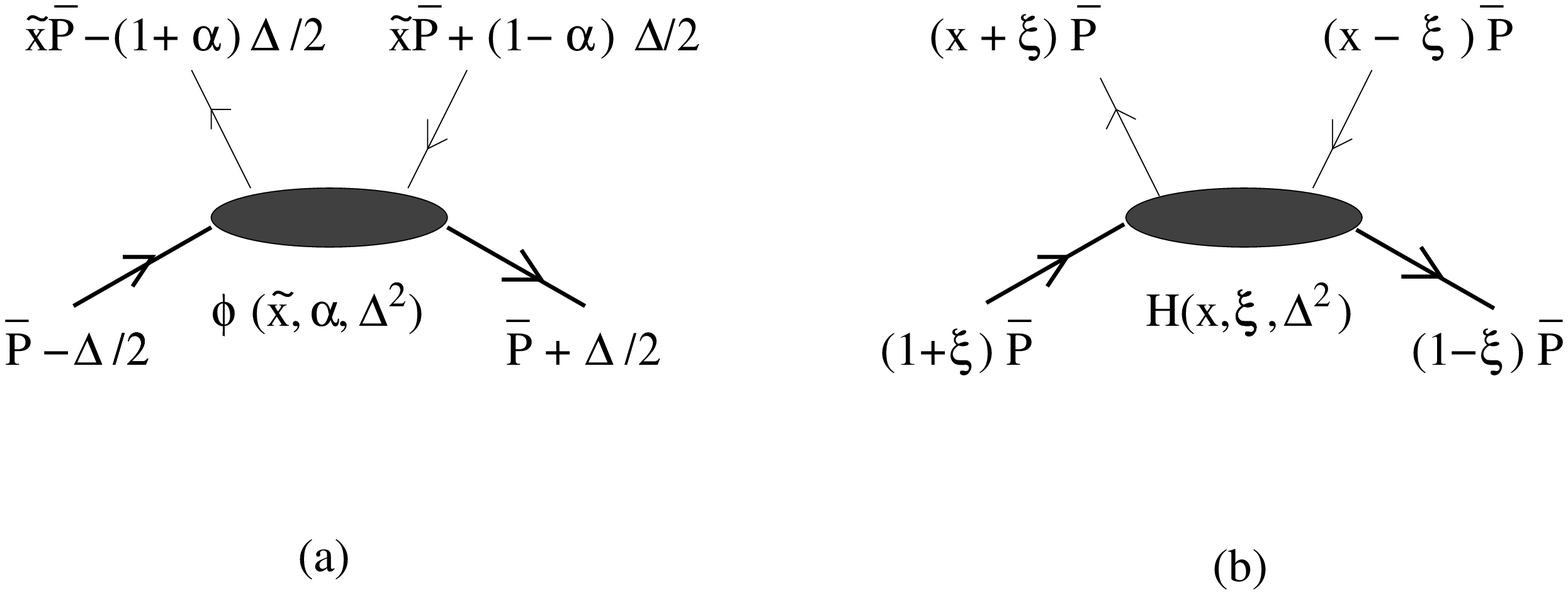}
\caption{}

\end{figure}

\newpage

\begin{figure}[h]
\vspace{10.0cm}
\includegraphics{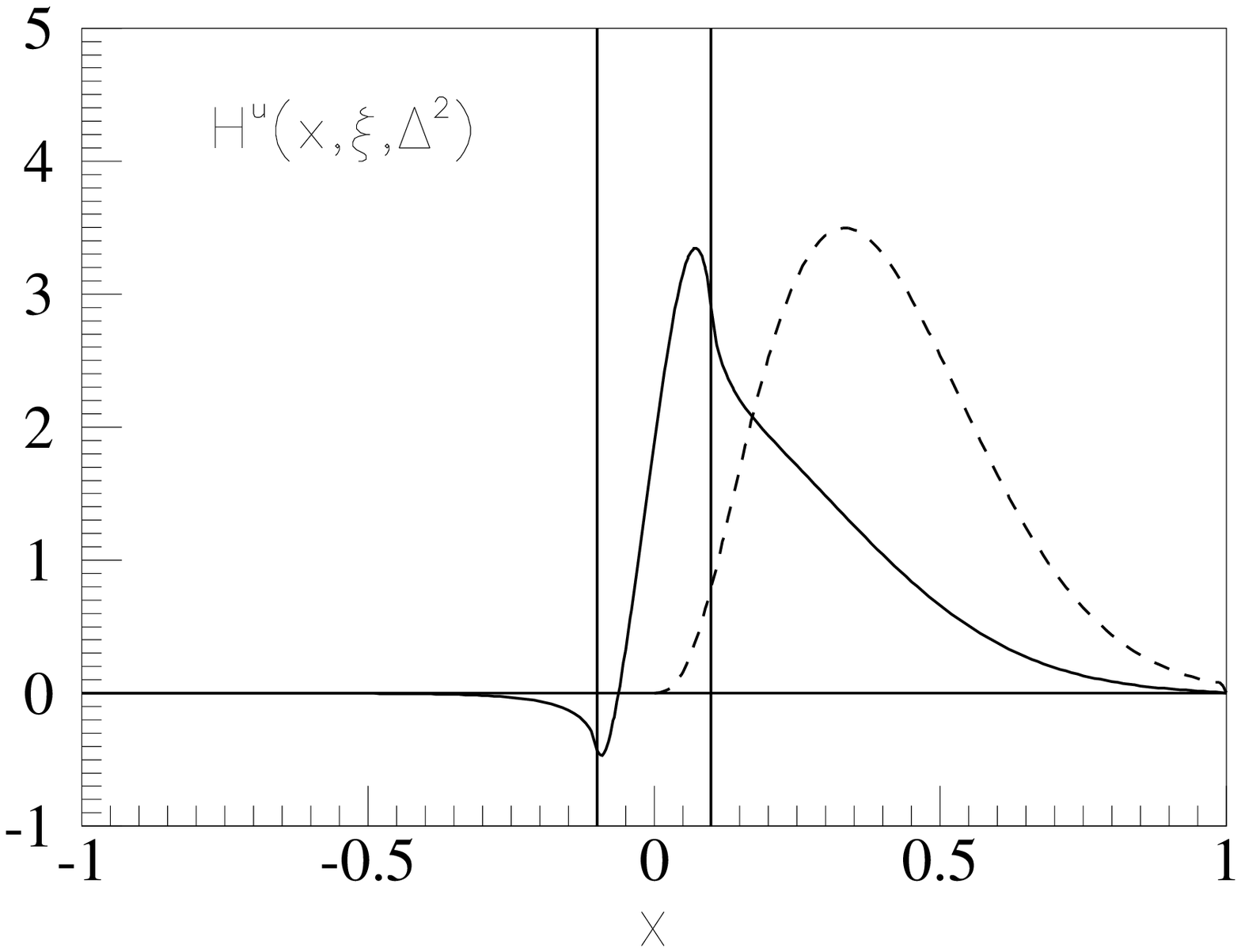}
\caption{}

\end{figure}

\newpage

\begin{figure}[h]
\vspace{14.cm}
\includegraphics{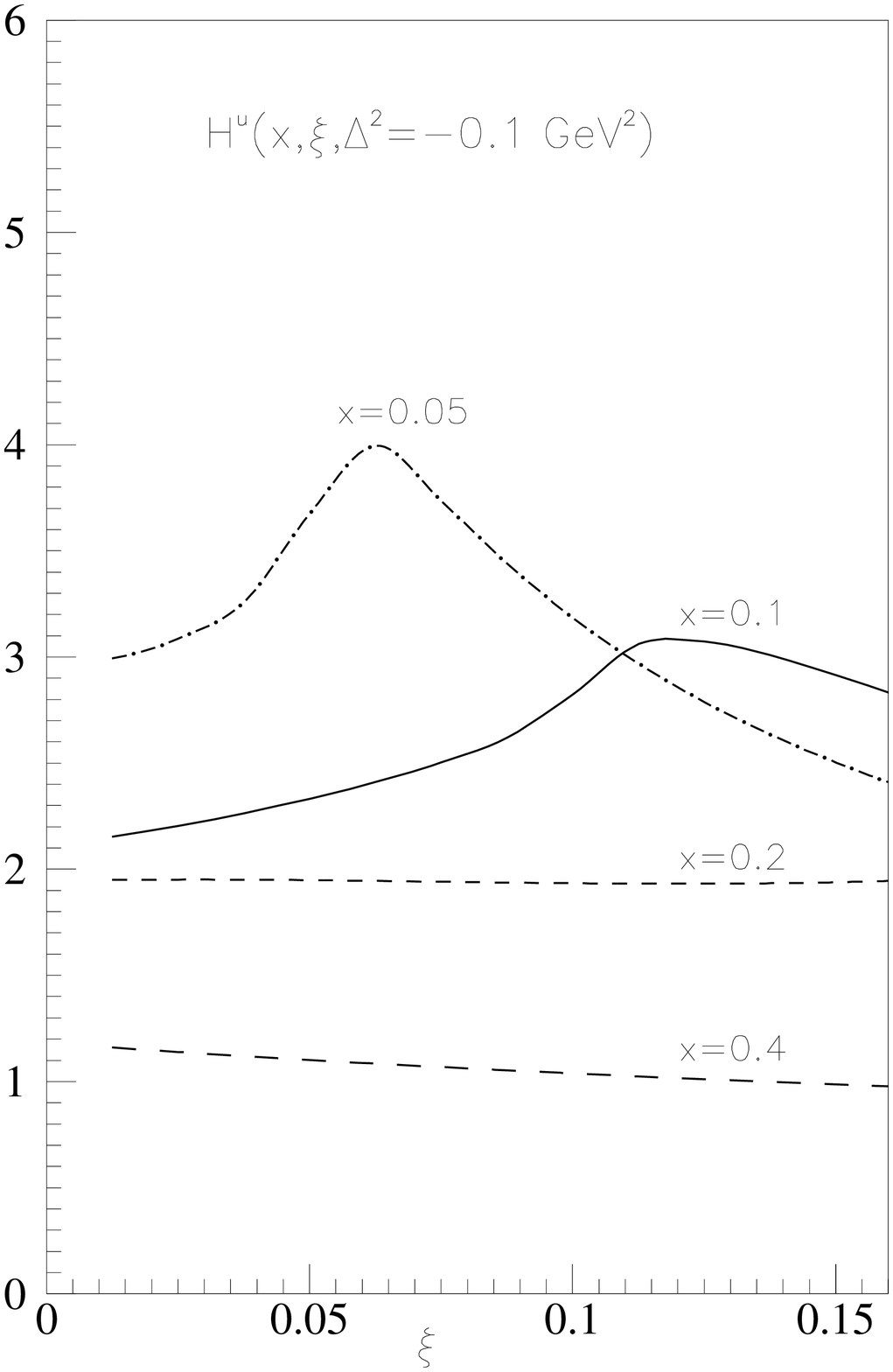}
\caption{}

\end{figure}

\newpage

\begin{figure}[h]
\vspace{12.cm}
\includegraphics{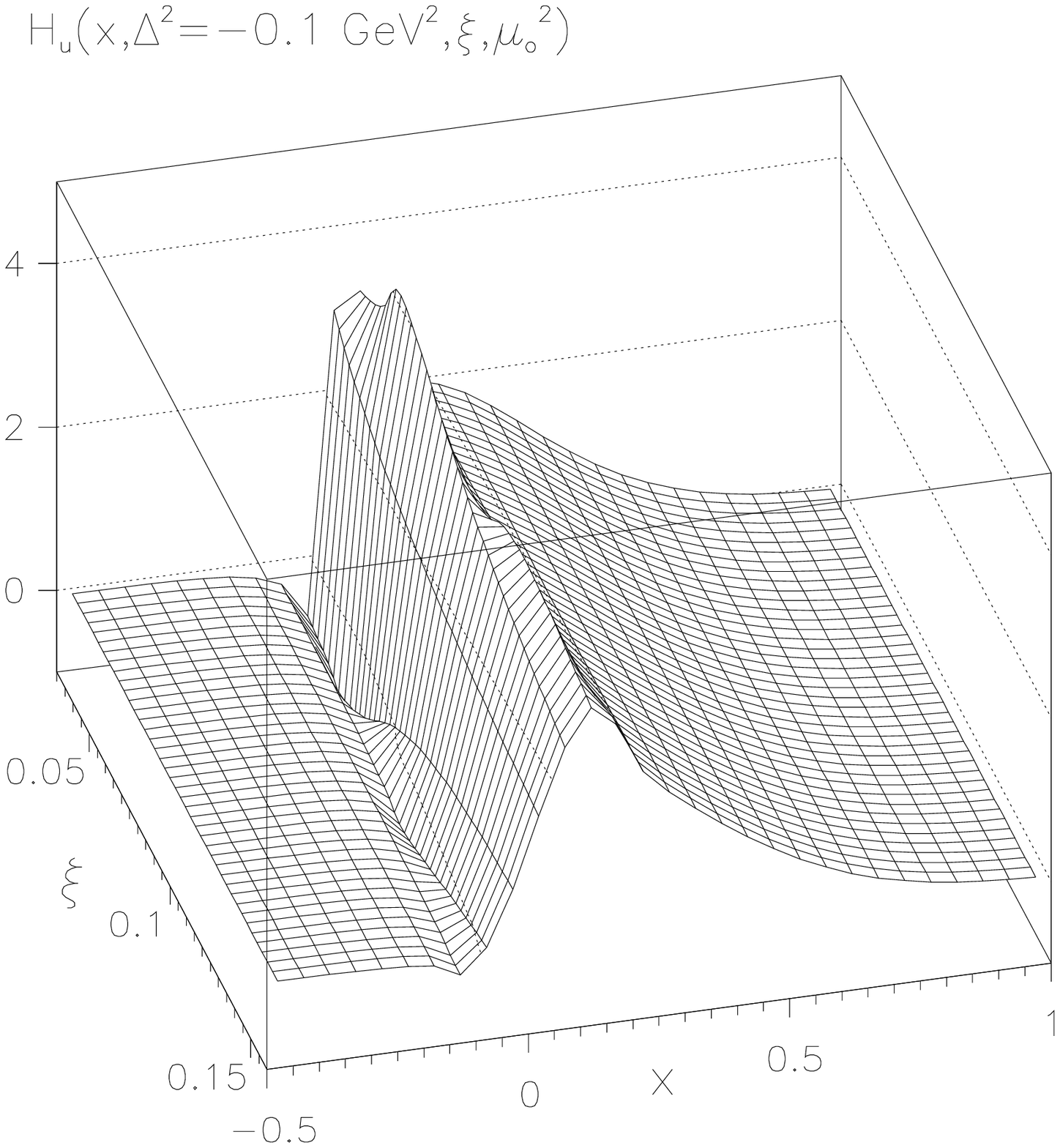}
\caption{}
\end{figure}

\newpage

\begin{figure}[h]
\vspace{8.2cm}
\includegraphics{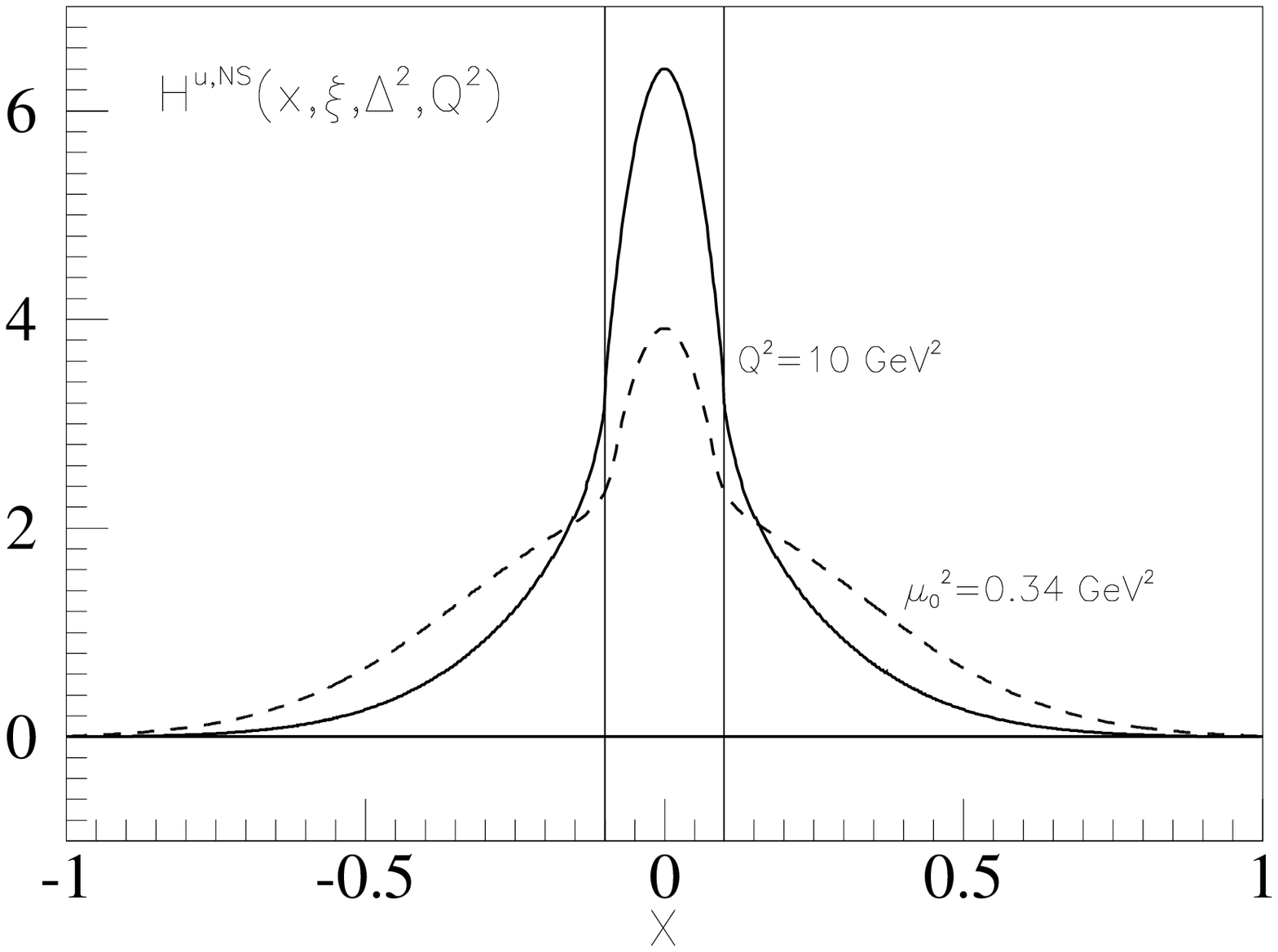}
\caption{}
\end{figure}

\vskip 3cm

\begin{figure}[h]
\vspace{8.2cm}
\includegraphics{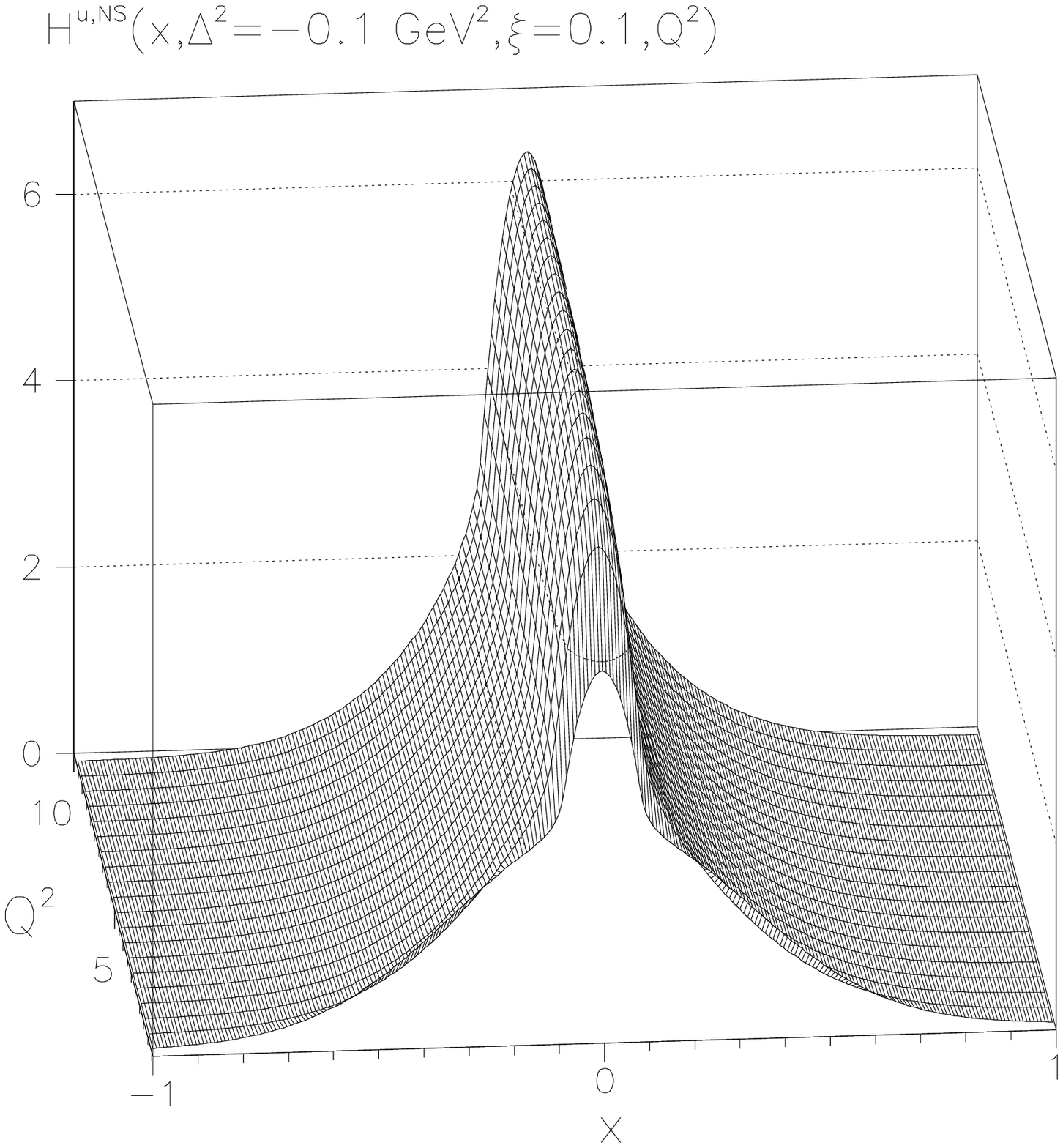}
\caption{}
\end{figure}

\newpage

\begin{figure}[h]
\vspace{9.2cm}
\includegraphics{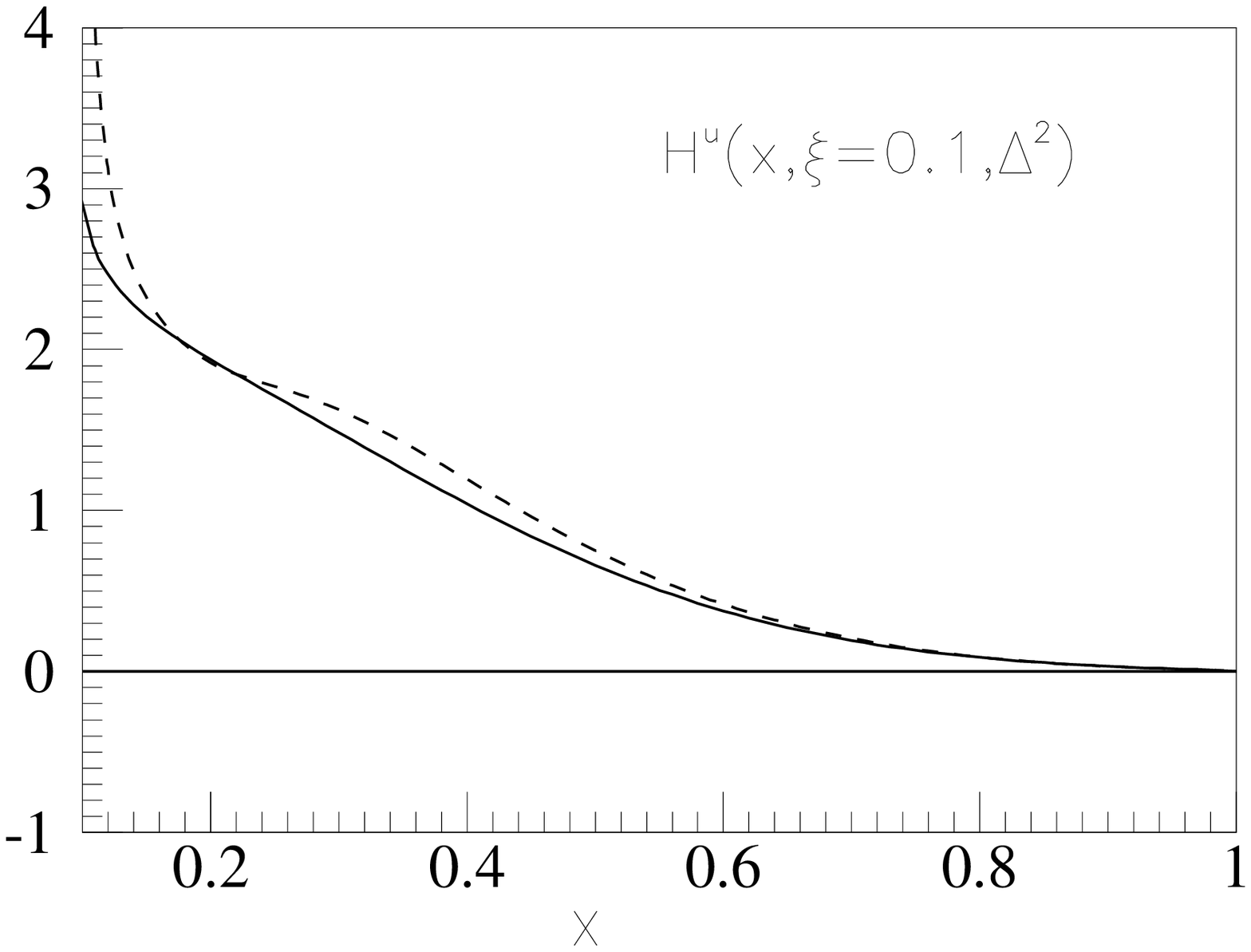}
\caption{}
\end{figure}

\end{document}